\documentclass[preprint,nofootinbib,aps,superscriptaddress,eqsecnum]{revtex4-1}
 \pdfoutput=1
\usepackage{amsmath,amssymb,graphicx,subfigure}

\newcommand{\beq}{\begin{equation}}
\newcommand{\eeq}{\end{equation}}
\newcommand{\bea}{\begin{eqnarray}}
\newcommand{\eea}{\end{eqnarray}}
\newcommand{\Z}{\mathbb{Z}}

\def\nn{\nonumber}

\def\ra{\rightarrow}

\def \met  {\mbox{${E\!\!\!\!/_T}$}}
\newcommand{\newc}{\newcommand}
\newc{\wt}{\widetilde}
\newc{\delhpm}{\Delta{M_{H^{\pm}}}}
\newc{\delh}{\Delta{M_{H}}}
\newc{\delhpmh}{\Delta{M_{H^{\pm}H}}}
\newc{\mdm}{M_A}
\newc{\mhpm}{M_{H^{\pm}}}
\newc{\mh}{M_{H}}
\newc{\lams}{\lambda_{S}}
\newc{\hp}{H^+}
\newc{\hm}{H^-}
\newc{\hpm}{H^{\pm}}
\newc{\hmp}{H^{\mp}}
\newc{\mwpm}{M_{W^{\pm}}}

\def\issue(#1,#2,#3){{\bf #1}, #2 (#3)}
\def\iss(#1,#2,#3){{\bf #1} (#3) #2}
\def\ASTR(#1,#2,#3){Astropart.\ Phys. \issue(#1,#2,#3)}
\def\AJ(#1,#2,#3){Astrophysical.\ Jour. \issue(#1,#2,#3)}
\def\AJS(#1,#2,#3){Astrophys.\ J.\ Suppl. \issue(#1,#2,#3)}
\def\APP(#1,#2,#3){Acta.\ Phys.\ Pol. \issue(#1,#2,#3)}
\def\JCAP(#1,#2,#3){Journal\ XX. \issue(#1,#2,#3)} 
\def\SC(#1,#2,#3){Science \issue(#1,#2,#3)}
\def\PRD(#1,#2,#3){Phys.\ Rev.\ D \issue(#1,#2,#3)}
\def\PR(#1,#2,#3){Phys.\ Rev.\ \issue(#1,#2,#3)} 
\def\PRC(#1,#2,#3){Phys.\ Rev.\ C \issue(#1,#2,#3)}
\def\NPB(#1,#2,#3){Nucl.\ Phys.\ B \issue(#1,#2,#3)}
\def\NPPS(#1,#2,#3){Nucl.\ Phys.\ Proc. \ Suppl \issue(#1,#2,#3)}
\def\NJP(#1,#2,#3){New.\ J.\ Phys. \issue(#1,#2,#3)}
\def\JP(#1,#2,#3){J.\ Phys.\issue(#1,#2,#3)}
\def\JPG(#1,#2,#3){J.\ Phys.\ G \issue(#1,#2,#3)}
\def\PL(#1,#2,#3){Phys.\ Lett. \issue(#1,#2,#3)}
\def\ZP(#1,#2,#3){Z.\ Phys. \issue(#1,#2,#3)}
\def\ZPC(#1,#2,#3){Z.\ Phys.\ C  \issue(#1,#2,#3)}
\def\PREP(#1,#2,#3){Phys.\ Rep. \issue(#1,#2,#3)}
\def\PRL(#1,#2,#3){Phys.\ Rev.\ Lett. \issue(#1,#2,#3)}
\def\MPL(#1,#2,#3){Mod.\ Phys.\ Lett. \issue(#1,#2,#3)}
\def\RMP(#1,#2,#3){Rev.\ Mod.\ Phys. \issue(#1,#2,#3)}
\def\SJNP(#1,#2,#3){Sov.\ J.\ Nucl.\ Phys. \issue(#1,#2,#3)}
\def\CPC(#1,#2,#3){Comp.\ Phys.\ Comm. \issue(#1,#2,#3)}
\def\IJMPA(#1,#2,#3){Int.\ J.\ Mod. \ Phys.\ A \issue(#1,#2,#3)}
\def\MPLA(#1,#2,#3){Mod.\ Phys.\ Lett.\ A \issue(#1,#2,#3)}
\def\PTP(#1,#2,#3){Prog.\ Theor.\ Phys. \issue(#1,#2,#3)}
\def\RMP(#1,#2,#3){Rev.\ Mod.\ Phys. \issue(#1,#2,#3)}
\def\NIMA(#1,#2,#3){Nucl.\ Instrum.\ Methods \ A \issue(#1,#2,#3)}
\def\EPJC(#1,#2,#3){Eur.\ Phys.\ J.\ C \issue(#1,#2,#3)}
\def\RPP (#1,#2,#3){Rept.\ Prog.\ Phys. \issue(#1,#2,#3)}
\def\PPNP(#1,#2,#3){ Prog.\ Part.\ Nucl.\ Phys. \issue(#1,#2,#3)}
\newc{\PRDR}[3]{{Phys. Rev. D} {\bf #1}, Rapid  Communications, #2 (#3)}

\def\PLB(#1,#2,#3){Phys.\ Lett.\ B  \issue(#1,#2,#3)}
\def\JHEP(#1,#2,#3){JHEP \issue(#1,#2,#3)}

\begin{document}
\title{Exploring collider signatures of the inert Higgs doublet model}

\author{Amitava Datta}\email{adatta{\_}ju@yahoo.co.in} 
\author{Nabanita Ganguly}\email{nabanita.rimpi@gmail.com} 
\affiliation{Department of Physics, University of Calcutta, 
92 Acharya Prafulla Chandra Road, Kolkata 700\,009, India}
\author{Najimuddin Khan}\email{phd11125102@iiti.ac.in} 
\author{Subhendu Rakshit}\email{rakshit@iiti.ac.in} 
\affiliation{Discipline of Physics, Indian Institute of Technology Indore,\\
 Khandwa Road, Simrol, Indore  453\,552, India \vspace{1.80cm}}	

\begin{abstract}\vspace*{10pt}

We revisit the multilepton ($ml$) + $\met$+ $X$ signatures of the Inert Doublet Model (IDM) of dark matter  in  future LHC experiments for $m =$ 3, 4 and simulate, for the first time, the $m = 5$ case. Here $X$ stands for any number of jets. We illustrate these signals with benchmark points consistent with the usual constraints like unitarity, perturbativity, the precision electroweak data, the observed dark matter relic density of the Universe and, most importantly,   the stringent LHC constraints from the post Higgs ($h$) discovery era like the measured $M_h$ and the upper bound on the invisible width of $h$ decay which were not included in earlier analyses of multilepton signatures. We find that if the IDM model is embedded in a grand dessert scenario so that the unitarity constraint holds up to a very high scale, the whole of the highly restricted parameter space allowed by the above constraints can be probed at the LHC via the $3l$ signal for an integrated luminosity $\sim 3000 ~{\rm fb}^{-1}$. On the other hand, if any new physics shows up at a scale $\sim$ 10~TeV, only a part of the enlarged allowed parameter space can be probed. The $4l$ and $5l$ signals can help to discriminate among different IDM scenarios as and when sufficient integrated luminosity accumulates.

\end{abstract}

\renewcommand{\thefootnote}{\arabic{footnote}}
\maketitle

\section{Introduction}
Discovery of a scalar boson~\cite{Aad:2012tfa,Chatrchyan:2012ufa} at the Large Hadron Collider (LHC) in 2012 with properties very similar to the Higgs boson ($h$) responsible for electroweak symmetry breaking in the Standard Model (SM) with a minimal scalar sector has validated this model. The mass of the Higgs like boson has been measured to be about 125 GeV~\cite{Aad:2012tfa,Chatrchyan:2012ufa,Giardino:2013bma}. So far, the LHC or other experiments have not discovered any signature of new physics beyond the SM. However, the non-zero neutrino masses, the baryon-antibaryon asymmetry in Nature, the presence of dark matter (DM) and dark energy in the Universe and many other observables compel us to look beyond the minimal SM. In this paper our focus will be on a popular model which can potentially explain the measured DM relic density in the Universe~\cite{Baer:2008uu}.

Strong astrophysical evidences suggest that our Universe is pervaded by DM. The relic density of DM is $\Omega h^2=0.1198\pm0.0026$ as measured by the satellite based experiments Planck~\cite{Ade:2013zuv} and WMAP~\cite{Bennett:2012zja} that are geared to the measurement of various properties of the cosmic microwave background radiation~(CMBR) with an unmatched precision. It will be doubly assuring to confirm the presence of DM by terrestrial experiments.
Many such experiments have been carried out for direct detection of DM via its scattering with the nucleons~\cite{Akerib:2013tjd,Aprile:2011hi,Aprile:2012nq}. However, no signal has been detected. Recent null results by the LUX experiments~\cite{Akerib:2013tjd} could eliminate a significant portion of the parameter space in the DM mass versus DM-nucleon cross-section plane. However, these constraints are marred by the uncertainties stemming from the assumption that the Earth is flying through a uniform DM cloud of significant density.
The clumpy nature of DM leaves open the possibility that the density of DM in the cosmologically tiny region surrounding the Earth, which has not been directly measured so far, is very small.
This makes the option that the DM may be produced directly at a high energy collider like LHC even more attractive. Weakly interacting massive particles~(WIMP) can indeed be produced by the proton--proton collisions at LHC which escape the detector leading to the celebrated missing energy signal.
As backgrounds are somewhat better understood in a man-made laboratory, it is not unreasonable to argue that a collider might be the best bet in revealing the true nature of DM particles.

The search for DM at the LHC is a topic of great contemporary interest. A large number of models compatible with the relic density data have been proposed and their prospective signatures at the LHC have been studied (see, e.g.,~\cite{Abercrombie:2015wmb,{Abdallah}}). The discovery of the Higgs boson~\cite{Aad:2012tfa,Chatrchyan:2012ufa,Giardino:2013bma} has completed the spectrum of the minimal version of the SM. Yet it must be admitted that the scalar sector of the SM is the least constrained one~\cite{CMS:2015kwa}. It is, therefore, quite probable that the DM particle has its abode in the extended scalar sector. A simple possibility, which we pursue in this paper, is to extend the scalar sector of the standard model~(SM) by a SU(2) doublet protected by a $\Z_2$ discrete symmetry. This model, known as the Inert Doublet Model, was first proposed by Deshpande and Ma~\cite{Deshpande:1977rw}. In the IDM heavier neutral and charged scalars do exist but do not take part in electroweak symmetry breaking. The exact $\Z_2$ symmetry does not allow the heavier neutral scalar to mix with the SM Higgs and as a result, it does not acquire a vacuum expectation value. In this model, the lightest scalar, odd under $\Z_2$, provides the WIMP DM candidate. This particle may be produced in association with a heavier scalar. It may also appear in the decay cascades of the heavier scalars which are produced in pairs. Both the processes yield the generic $m$-leptons + $n$-jets + $\met$ signatures, where the lepton and jets come mainly from $W^\pm$ and $Z$ bosons which also appear in the above decay cascades. For $m=0$ the signal is relatively large but this electroweak jet production
is easily swamped by the huge QCD background. As has already been noted in the literature and will be reiterated in this paper, the $\met$ in the signal is rather modest which is not enough to discriminate against the strong QCD backgrounds with cross sections several orders of magnitude larger. It should be borne in mind that in a hadron collider the latter processes also involve a sizeable $\met$ due to mismeasurement of jet energies, underlying events etc. The main attention has, therefore, been focused on multilepton + $\met$ signatures. Here one has to contest the electroweak backgrounds with a relatively small cross section. The task, nevertheless, is uphill as both the signal and the background, which typically has a much larger size, involve leptons coming from $W^\pm$ and $Z$ decays. However, after adjusting the cuts a modest signal to background ratio can be salvaged at the LHC experiments with upgraded luminosity especially if the signal involves virtual Zs. This will be shown below.

The most well studied signal of the DM-Higgs coupling, both phenomenologically~\cite{Battaglia:2004js,Drozd:2011aa,
Gopalakrishna:2009yz,Ghosh:2012ep} and experimentally~\cite{Aad:2015txa,CMS:2016jjx,Chatrchyan:2014tja,Trocino:2016zde}, has been the invisible decay of $h$. This occurs provided the mass of the DM particle is $< M_h/2$.  However, since this generic signature may arise in any model where $h$ can decay into a pair of long lived WIMPs not necessarily the DM particle,
it is hard to figure out the underlying physics form this signal alone. The next simplest case is the dilepton + $\met$ topology ($m=2$). This has already been studied in the context of the IDM~\cite{Dolle:2009ft,Belanger:2015kga}. It has been noted that the LHC Run 1 data in this channel is sensitive only to regions of the parameter space  not containing a viable DM candidate~\cite{Belanger:2015kga}. Nevertheless the authors optimistically expected an observable signal during Run 2~\cite{Belanger:2015kga}. It should, however, be stressed that even if both these signals show up it will still be difficult to reveal the new physics involved. Additional search channels, therefore, are always welcome.

Signatures with $m = 3, 4$ were studied in the  IDM~\cite{ Miao:2010rg,Gustafsson:2012aj}. However,  the  parameter space of the IDM is constrained by a plethora of important  constraints both theoretical and empirical (see  section 3 for references and further details). 
In \cite{ Miao:2010rg,Gustafsson:2012aj} the above signals were illustrated 
with benchmark points (BPs) not compatible with some important LHC constraints in the post Higgs discovery era like the measurement of the Higgs boson mass, the strong upper bound on the invisible decay width of the Higgs boson \cite{Belanger:2013xza}.  The main emphasis of this paper is to assess the prospect of these signals with a new set of realistic BPs consistent with more recent and stronger constraints. More important, we have not restricted our analyses to isolated BPs only. We have identified, as and when possible, the portions of the allowed parameter spaces (APS) of several representative scenarios sensitive to the proposed  signals in future LHC experiments. Then we have illustrated the features of the prospective signals with the help of BPs. Finally we have studied, for the first time,  the  $5l +\met$ signal.

Unlike some of the earlier analyses we do not impose any jet veto on the multilepton final states. Generically therefore, the signatures studied have the topology $m$-leptons + $\met$ + $X$, where $X$ stands for any number of jets. This choice is necessitated by the fact that the leptonic final states arising from the decays of the extra scalars in the IDM are often accompanied by ISR jets and a good fraction of the signal may be lost if the jet veto is imposed. This strategy is similar to the ones currently adopted by the LHC collaborations for multilepton analyses.

The plan of the paper is as follows. In section 2 we have briefly reviewed the salient features of the IDM. In the 
next section we have introduced three representative scenarios in the IDM and studied the APS in each case in the light of the available constraints. In section 4 we have studied the portions of the above APSs which are within the reach of future LHC experiments. Illustrative numerical results in each scenario are provided  with the help of several BPs. The main conclusions are summarized in the last section.

\section{Inert Doublet Model}
\label{sec:IDM}
In this model, the standard model is extended by adding an extra $SU(2)$ doublet scalar, odd under an additional discrete $\Z_2$ symmetry. Under this symmetry, all standard model fields are even. The $\Z_2$ symmetry prohibits the inert doublet to acquire a vacuum expectation value.

The renormalizable $CP$-conserving scalar potential at the tree level is given by~\cite{Deshpande:1977rw}
\bea
V(\Phi_1,\Phi_2) &=& \mu_1^2 |\Phi_1|^2 + \lambda_1 |\Phi_1|^4+ \mu_2^2 |\Phi_2|^2 + \lambda_2 |\Phi_2|^4\nn\\
&&+\: \lambda_3 |\Phi_1|^2 |\Phi_2|^2 
+  \lambda_4 |\Phi_1^\dagger \Phi_2|^2 + \frac{\lambda_5}{2} \left[ (\Phi_1^\dagger \Phi_2)^2 + {\rm h.c.}\right]  \, ,
\label{Scalarpot}
\eea
where $\mu_{1,2}$ and $\lambda_{i}$ ($i=1,2,3,4,5$) are real parameters. The SM Higgs doublet $\Phi_1$ and the inert doublet $\Phi_2$ are given by,
\beq
	\Phi_1 ~=~ \left(\begin{array}{c} G^+ \\ \frac{1}{\sqrt{2}}\left(v+h+i G^0\right) \end{array} \right),
	\qquad
	\Phi_2 ~=~ \left(\begin{array}{c} H^+\\ \frac{1}{\sqrt{2}}\left(H+i A\right) \end{array} \right) \, \nn 
\eeq
where, $v=246.221$ GeV is the vacuum expectation value of the $\Phi_1$, $G^\pm$ and $G^0$ are Goldstone bosons and $h$ is the SM Higgs. 

$\Phi_2$ contains a $CP$ even neutral scalar $H$, a $CP$ odd neutral scalar $A$, and a pair of charged scalar fields $H^\pm$.
The $\Z_2$ symmetry prohibits an $odd$ number of these scalar fields couple with the SM particles.
Either of the lightest neutral components $H$ and $A$ is stable and may be considered as a DM candidate.

After electroweak symmetry breaking, the scalar potential is given by,
\bea
V(h, H,A,H^\pm) &=&  \frac{1}{4} \left[ 2 \mu_1^2 (h+v)^2 + \lambda_1 (h+v)^4 +2 \mu_2^2 (A^2+H^2+2 H^+ H^-) \right. \nn \\
&& \left. + \lambda_2 (A^2 + H^2 + 2 H^+ H^-)^2  \right] \nn \\
&& + \frac{1}{2} (h+v)^2 \left[  \lambda_3 H^+ H^- 
+  \lambda_S  A^2  
+  \lambda_L  H^2 \right] \label{Scalarpot2}
\eea
where, 
\bea
\lambda_{L,S}&=&\frac{1}{2}\left(\lambda_3+\lambda_4\pm\lambda_5\right) \, .
\eea
Masses of these scalars are given by,
\begin{align}
	M_{h}^2 &= \mu_1^2 + 3 \lambda_1 v^2,\nn \\
	M_{H}^2 &= \mu_2^2 +  \lambda_L v^2, \nn \\
	M_{A}^2 &= \mu_2^2 + \lambda_S v^2,\nn \\
	M_{H^\pm}^2 &= \mu_2^2 + \frac{1}{2} \lambda_3 v^2  \,\nn .
\end{align}

For $\lambda_4-\lambda_5<0$ and $\lambda_5>0$~($\lambda_4+\lambda_5<0$ and $\lambda_5<0$), 
$A$~($H$) is the lightest $\Z_2$ odd particle (LOP). In this work, we take $A$ as the LOP and hence, as a viable DM candidate. Choice of $H$ as LOP will lead to similar results. 

For analyses in the next two section we define
\bea
	\Delta M_H&\equiv& M_H-M_A, \nn\\
	\Delta M_{H^\pm}&\equiv&M_{H^\pm}-M_A \nn\, .
\eea
so that the independent parameters for the IDM become $\{M_A, \Delta M_H, \Delta M_{H^\pm}, \lambda_2, \lambda_S\}$. Here we have chosen the Higgs portal coupling $\lambda_S$ as we treat $A$ as the DM particle.
Moreover $\lambda_2$ does not play any role in relic density calculation.  
Nor does it directly affect the masses of the inert scalars
which determine the collider signatures~\cite{Belanger:2015kga}.
We have, therefore,  chosen  $\lambda_2 = 0$\footnote{ However, 
it had been shown in Refs.~\cite{Ginzburg:2010wa,Sokolowska:2011yi,Sokolowska:2011sb,Sokolowska:2011aa,Krawczyk:2013wya,Ginzburg:2009dp} that depending
on the parameter space, this choice  might lead to a $\Z_2$ violating vacuum at
finite temperatures. During the thermal evolution of the Universe, if
the Universe happens to rest in such a vacuum for long, it might lead
to  intriguing cosmological implications~\cite{Ginzburg:2009dp}. A detailed study of this parameter space dependent  effect is beyond the scope of this paper.}.

\section{The constraints on the IDM and their implications}
\label{sec:constraints}
The five dimensional parameter space of this model discussed in the last section is constrained by various theoretical considerations like stability of the vacuum, perturbativity and unitarity of the scattering matrix. Experimental constraints such as the electroweak precision measurements, the direct search limits from LEP and the Higgs invisible decay width measured at the LHC also impose additional constraints. Last but not the least the requirement that the IDM alone saturates the measured DM relic density of the Universe is also instrumental in obtaining a finite APS to be tested at the LHC. Recently detailed  bounds on the IDM have been studied by several groups, see e.g., Refs.~\cite{LopezHonorez:2006gr,LopezHonorez:2010tb,Hambye:2009pw,
Arhrib:2013ela,Modak:2015uda,Gustafsson:2010zz,
Hashemi:2015swh,Aoki:2013lhm,Ferreira:2015pfi,Ferreira:2015dma,
Ilnicka:2015sra,Swiezewska:2015paa,
Krawczyk:2013pea,Gorczyca:2011rs, Khan:2015ipa,Ilnicka:2015jba,Chakrabarty:2015yia,Belyaev:2016lok}.

In this study we first determine the APS consistent with the above constraints 
for three representative scenarios: 
\begin{itemize}
\item[] {\bf A)}  $M_A = 70.0$ GeV, $\lambda_{S}$ = 0.005 (Fig.~\ref{fig:STUcheck} of this paper).
\item[] {\bf B)} $M_A = 70.0$ GeV, $\lambda_{S}$ = 0.007 (Fig.~2 of Ref. \cite{Khan:2015ipa}).
\item[] {\bf C)} $M_A = 55.0$ GeV, $\lambda_{S}$ = 0.0035 (Fig.~\ref{fig:STUcheck55} of this paper).
\end{itemize}
Comparison of A) and B) highlights
the changes in the APS with $\lambda_{S}$ for fixed $M_A$. On the other hand C) represents a parameter space where invisible Higgs decay is allowed. In all three cases the free parameters $\Delta M_H$ and $\Delta M_{H^\pm}$ delineate the APS restricted by the above constraints. These parameters along with $M_A$ govern the prospective LHC signatures in each scenario to be studied in the following. It may be noted that in scenario C) the constraint from the invisible decay width of $h$ requires $\lambda_{S}$ to be $\sim 10^{-3}$. However in A) and B) larger $\lambda_{S}$ could have been chosen as long as the choice was consistent with the observed DM relic density \cite{Ade:2013zuv,Bennett:2012zja}. Of course much larger $\lambda_{S}$ would be in conflict with the bounds from direct DM detection experiments \cite{Akerib:2013tjd}.
However, as argued in the introduction, these bounds are not yet compelling. Nevertheless we have restricted ourselves to choices consistent with the LUX data. In any case larger values of $\lambda_{S}$ does not change our main results qualitatively. 

\subsection{Vacuum stability bounds}
The stability of the scalar potential requires that the potential should not be
unbounded from below, i.e, it should not approach negative infinity along any direction in the field space at large field values.
The tree level scalar potential potential $V(\Phi_1,\Phi_2)$ is stable and bounded from below if~\cite{Deshpande:1977rw} 
\beq
\lambda_{1,2}(\Lambda) \geq 0, \quad \lambda_{3}(\Lambda) \geq -2 \sqrt{ \lambda_{1}(\Lambda)\lambda_{2}(\Lambda)}, \quad \lambda_{L,S}(\Lambda) \geq - \sqrt{ \lambda_{1}(\Lambda)\lambda_{2}(\Lambda)} \label{stabilitybound}
\eeq
where the coupling constants are evaluated at a scale $\Lambda$ using RG equations. However, for the parameter spaces considered in this paper the consequences of this bound are covered by other constraints. 
\subsection{Perturbativity bounds}

For the IDM to behave as a perturbative quantum field theory at any given scale $\Lambda$, one must impose the conditions on the couplings of radiatively improved scalar potential $V(\Phi_1,\Phi_2)$ as,
\beq
\mid \lambda_{1,2,3,4,5}\mid \leq 4 \pi \, .
\eeq
These upper bounds on the couplings $\lambda_i$'s at $\Lambda$  restrict $\Delta M_H$ and $\Delta M_{H^\pm}$. 
\subsection{Unitarity bounds}
The parameters of the scalar potential are severely constrained by the unitarity of the S-matrix, which at high energies consists of the quartic couplings $\lambda_i$'s of the scalar potential.
At very high field values, one obtains the S-matrix by using various scalar-scalar, gauge boson-gauge boson, and scalar-gauge boson scatterings~\cite{Lee:1977eg}.
The unitarity of the S-matrix demand that the absolute eigenvalues of the scattering matrix should be less than $8\pi$~\cite{Arhrib:2012ia,Grinstein:2015rtl,Cacchio:2016qyh} upto a certain scale $\Lambda$.
In this analysis we consider two choices of $\Lambda$: (i) the Planck scale and (ii) 10 TeV.
The former choice, representing the case where the IDM is the low energy sector of grand dessert model, imposes very strong constraints on the allowed region in the $\Delta M_H-\Delta M_{H^\pm}$-plane as can be seen from the bounded regions in the lower left corners of Fig.~\ref{fig:STUcheck} (scenario A)) and Fig.~\ref{fig:STUcheck55} (scenario C)) of this paper as well as Figs. 2 (scenario B)) and 5 of Ref.~\cite{Khan:2015ipa}\footnote{We caution the reader that the color conventions are not the same in different figures.}.
It also follows by comparing these figures that for small $\lambda_S$, this constraint is fairly insensitive to the choice of of $\lambda_S$. 
For the latter choice of the scale signifies the onset of some beyond IDM physics at a scale $\sim$ 10 TeV. Here the relaxed unitarity constraints are much weaker leading to a larger APS in each case as is illustrated by the light green region of Fig.~\ref{fig:STUcheck} and the blue region in Fig.~\ref{fig:STUcheck55}\footnote{The narrow gaps beyond the left and the lower edge of the blue region in Fig.~\ref{fig:STUcheck55} are due to the DM constraint as we shall see below.}. We have checked that the entire parameter space shown in Fig.~2 of Ref.~\cite{Khan:2015ipa} is allowed by the relaxed unitarity constraint.

\subsection{Bounds from electroweak precision experiments}
\begin{figure}[ht]
 \begin{center}
  {
\includegraphics[width=2.8in,height=2.8in, angle=0]{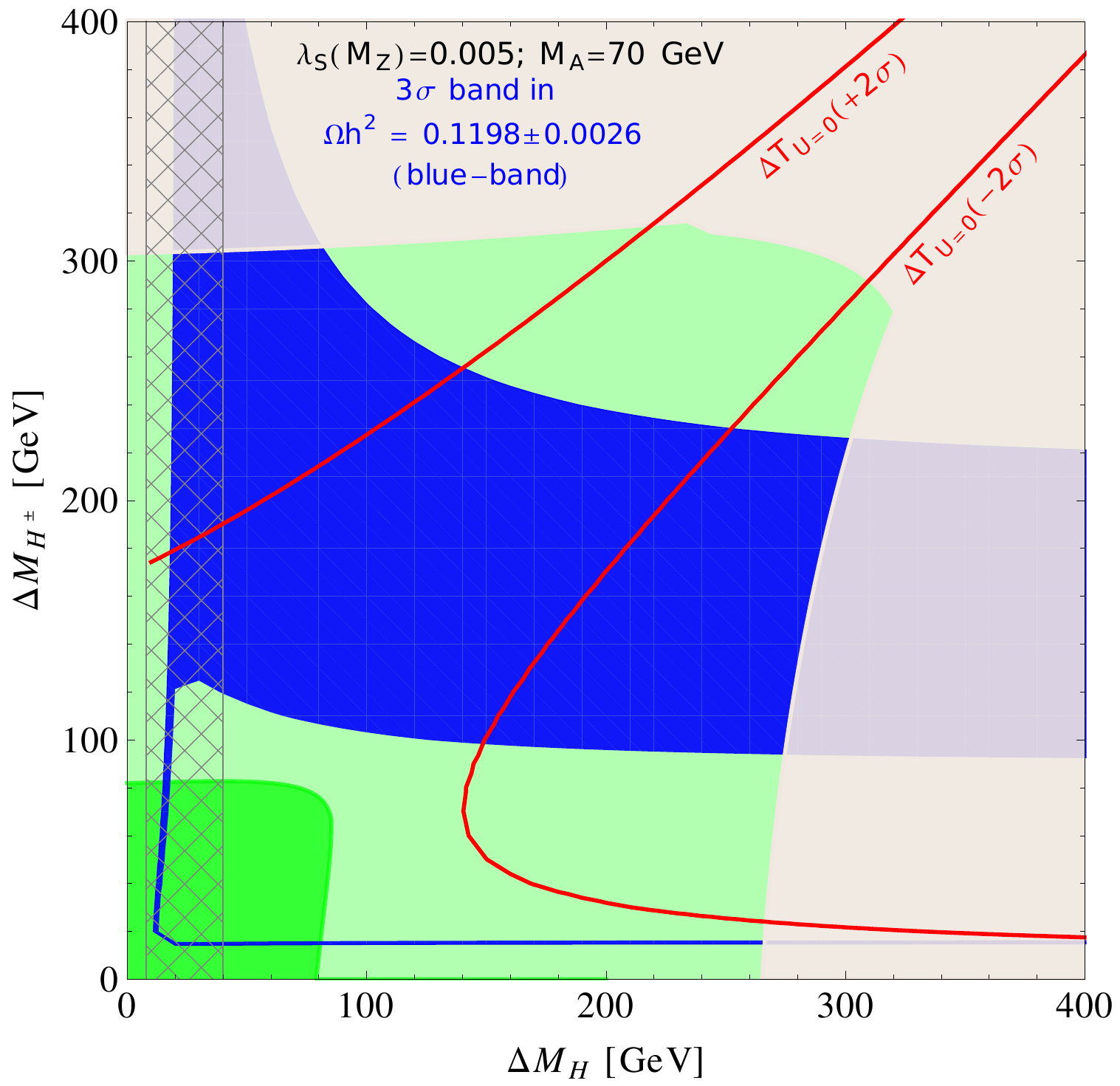}}
  \caption{\label{fig:STUcheck} \textit{\rm{The allowed parameter space in $\Delta M_{H}-\Delta M_{H^\pm}$ plane for $M_A=70$~GeV and $\lambda_S=0.005$. The constraints from the $T$ parameter allows only the area between the solid red lines. In the green region in the lower left corner the unitarity bound is valid upto the Planck scale. In the light green region the unitarity bound is valid upto 10 ${\rm TeV}$. The blue regions are allowed by the DM constraint at the 3$\sigma$ level \cite{Ade:2013zuv} and the relaxed unitarity constraint. The cross-hatched region is excluded from LEP\,II data.}}}
 \end{center}
 \end{figure}

Electroweak precision data has imposed bounds on the IDM via the Peskin-Takeuchi~\cite{Peskin:1991sw} 
parameters $S, ~T, ~U$ and the contributions of the additional scalars of the 
IDM to these parameters can be found in Refs.~\cite{Barbieri:2006dq,Arhrib:2012ia}.
We use the NNLO global electroweak fit results obtained by the Gfitter group~\cite{Baak:2014ora}, 
\beq
\Delta S = 0.06 \pm 0.09, ~ ~ \Delta T = 0.1 \pm 0.07
\label{STU1}
\eeq
with a correlation coefficient of $+0.91$, for $\Delta U$ to be zero. We use eqn.~\ref{STU1} as the contributions of the scalars in the IDM to $\Delta U$ are indeed negligible. 
In Fig.~\ref{fig:STUcheck}, Fig.~\ref{fig:STUcheck55} and Fig. 2 of Ref.~\cite{Khan:2015ipa}, the parameter space allowed by the $\Delta T$ constraint at the $2\sigma$
level is the region between the two red solid curves (the same color convention has been used in all figures). This constraint is roughly independent of $\lambda_S$. The entire parameter space in Fig.~\ref{fig:STUcheck} is allowed by the $\Delta S$ constraint at the $2\sigma$ level. It is well known that the $\Delta T$ constraint
primarily restricts the $SU(2)$ breaking parameter $\Delta M_H$ and $\Delta M_{H^\pm}$. However, it should be borne in mind that large values of these mass differences, though allowed by the electroweak precision data, are forbidden by the perturbativity and unitarity constraints. This results in an APS bounded from above as is illustrated in the above figures.
\subsection{Bounds from LHC diphoton signal strength}
In the IDM, the $H^\pm$ gives additional contributions to diphoton decay of the Higgs at one loop. 
The analytical expressions can be found in Refs.~\cite{Djouadi:2005gj, Swiezewska:2012eh,Krawczyk:2013jta}.
The measured values are $\mu_{\gamma\gamma}=1.17\pm0.27$ from ATLAS~\cite{Aad:2014eha} and $\mu_{\gamma\gamma}=1.14^{+0.26}_{-0.23}$ from CMS~\cite{Khachatryan:2014ira}. 
The benchmark points used in this paper, are allowed at 1.5$\sigma$ by both the experiments.
\subsection{Invisible Higgs decay bounds form LHC}
If the inert particle mass less than $\frac{M_h}{2}$ then Higgs can decays to pair of inert particles.
LHC invisible Higgs decay~\cite{Aad:2015txa,CMS:2016jjx,Chatrchyan:2014tja} width puts stringent constraints on the parameter spaces for inert particle mass less than $\frac{M_h}{2}$. For more details see Refs.~\cite{Arhrib:2013ela, Khan:2015ipa,Goudelis:2013uca}. In scenario C) the BR of invisible $h$ decay is approximately 0.05. 
In Fig.~\ref{fig:relicvsmassIn}, we illustrate the parameter space where relic density is in the right ballpark as a function of DM masses for three choices of $\lambda_s$. 

\begin{figure}[ht]
 \begin{center}
  {
\includegraphics[width=2.8in,height=2.8in, angle=0]{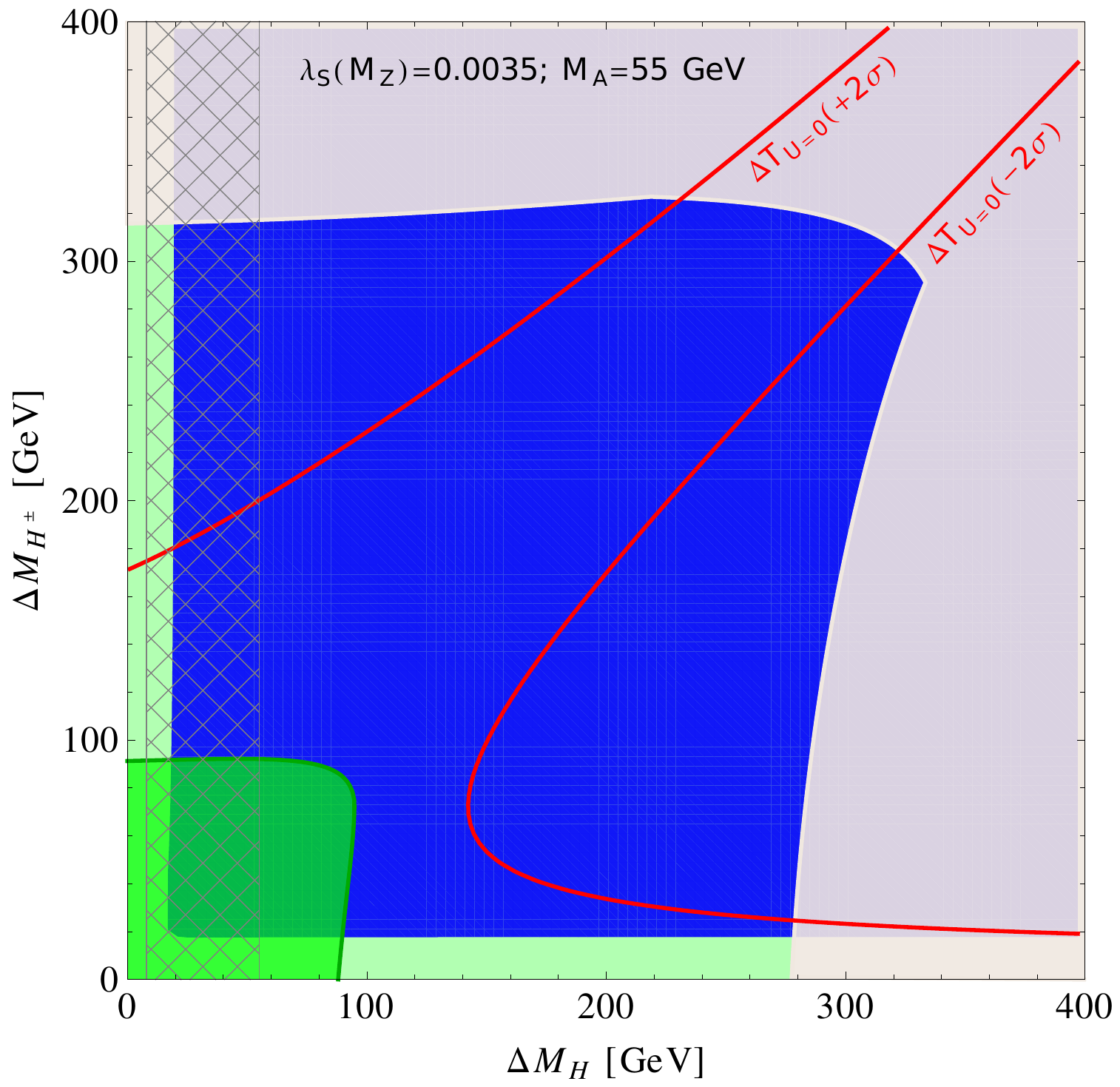}}
  \caption{\label{fig:STUcheck55} \textit{\rm{The allowed parameter space in $\Delta M_{H}-\Delta M_{H^\pm}$ plane for $M_A=55$~GeV and $\lambda_S=0.0035$. The constraints from the $T$ parameter, LEP\,II and the stronger unitarity condition are as in Fig. \ref{fig:STUcheck}. The blue region is allowed by the DM constraint at the 3$\sigma$ level \cite{Ade:2013zuv} and the weaker unitarity condition.}} } 
 \end{center}
 \end{figure} 
\subsection{Direct search limits from LEP}

The decays $Z\ra AH$, $Z\ra H^+ H^-$, $W^\pm\ra A H^\pm$, and $W^\pm\ra H H^\pm$ are restricted from $Z$ and $W^\pm$ decay widths at LEP.
It implies $M_{A} + M_{H} \geq M_{Z}$, $2M_{H^\pm}\geq M_{Z}$, and $M_{H^\pm} + M_{H,A} \geq M_{W}$. More stringent constraints on the IDM can be extracted from chargino~\cite{Pierce:2007ut} and neutralino~\cite{Lundstrom:2008ai} searches at LEP\,II: The charged Higgs mass $M_{H^\pm}\geq 70$~GeV.
The bound on $M_A$ is rather involved: If $M_A<80$ GeV, then $\Delta M_H$ should be either $\leq 8$ GeV or $\gtrsim 110$~GeV ({see Fig. 7 of Ref.~\cite{Lundstrom:2008ai}).

\subsection{Constraints from dark matter relic density}
 \begin{figure}[ht]
 \begin{center}
 {
 \includegraphics[width=2.7in,height=2.62in, angle=0]{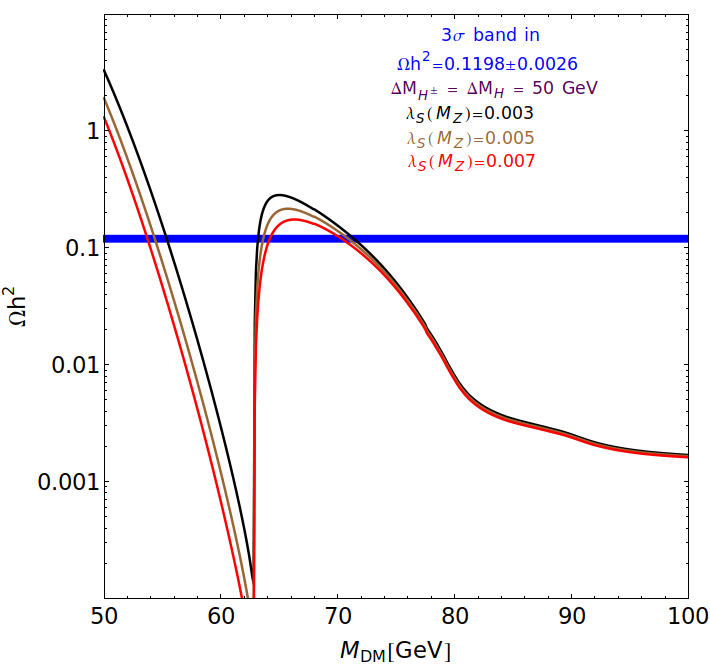}}
 \caption{\label{fig:relicvsmassIn} \textit{ The DM relic density $\Omega h^2$ as a function of the DM mass $M_{DM}(\equiv M_A)$ for different values of the coupling: $\lambda_S(M_Z)$ =0.003 (black), 0.005 (brown) and 0.007 (red), with $\Delta M_{H^\pm}=\Delta M_{H}=50$ GeV. The thin blue band corresponds to the observed DM relic density of the Universe at the $3\sigma$ level. 
 } }
 \end{center}
 \end{figure}
In the relic density calculation, the parameters $\{M_A, \Delta M_H, \Delta M_{H^\pm}, \lambda_S\}$ play pivotal roles.
In this model DM masses below 50 GeV are excluded by the measured DM relic density of the Universe and the invisible decay width~\cite{Cao:2007rm} of the Higgs from LHC global fit~\cite{Belanger:2013xza}. 
For $50 < M_A < 75$ GeV (see Fig.~\ref{fig:relicvsmassIn}) and $M_A \gtrsim 500$~GeV, we get DM relic density in the right ballpark.
However, we do not pursue the heavy $M_A$ option any further as it does not lead to interesting LHC signatures.
For further details we refer the reader to Refs.~\cite{Khan:2015ipa,Goudelis:2013uca}.
In Fig.~\ref{fig:STUcheck} (scenario A)), the upper and the narrow lower blue bands correspond to relic densities allowed by Planck~\cite{Ade:2013zuv} and WMAP~\cite{Bennett:2012zja} data within 3$\sigma$ for $M_A=70$ GeV and $\lambda_S=0.005$.
The dominant contribution comes from the process $AA\ra W^{\pm}W^{\mp*}$\footnote{The virtual $W^{\pm*}$ decays to quarks and leptons.},
although the process $AA\ra ZZ^{*}$ also contributes modestly. When $M_H$/$M_{H^\pm}$ is close to the DM mass (the narrow blue bands in Fig.~\ref{fig:STUcheck}), the contributions coming from co-annihilation~\cite{griest} between $A,H$/$A,H^\pm$ are also significant~\cite{Queiroz:2015utg}.
Due to the stronger unitarity constraint upto the Planck scale (see the darker green region of Fig.~\ref{fig:STUcheck}) only the thin lower blue band is allowed.
If instead the weaker unitarity constraint is imposed then a sizeable part of the upper blue band is also allowed.
Comparing with Fig. 2 of Ref.~\cite{Khan:2015ipa}, it follows that for a larger $\lambda_S$ ($ = 0.007$), the upper blue band shifts downwards along the $\Delta M_{H^\pm}$-axis.
In this case the enhanced contributions coming from $h$-mediated $s$-channel processes allow the lower values of $\Delta M_{H^\pm}$. However, $\Delta M_H$ almost remains unchanged. On the other hand for still smaller $\lambda_S$ 
(e.g., $ 0.001$) the DM constraint allow only narrow strips in the parameter space
(see Fig. 5 of Ref.~\cite{Khan:2015ipa}).
It is worth noting that the apparently large APS, which opens up due to the relaxation of unitarity constraints, is severely reduced by the very tight relic density constraints resulting in
 an APS bounded from above. In scenario C) (Fig. 2) 
the main contribution to the relic density comes from the process $A A \ra b \bar{b}$ via $h$ exchange. 
The parameter space allowed by the DM constraint in this case is much larger compared to the other scenarios. The entire $\delhpm-\delh$ plane for $\delhpm, \delh \gtrsim 12$ GeV is allowed by WMAP and Planck data. Of course the unitarity bounds and the $T$-parameter constraints restores a finite APS.

We conclude by noting that the allowed parameter space of the IDM is severely restricted by the perturbativity, unitarity, electroweak precision data and DM relic density
constraints resulting in a bounded APS in all the scenarios we have studied. In the next turn our attention to the LHC signatures viable in these APSs and to illustrate the signatures with benchmark points consistent with
all constraints.
\section{The multilepton signatures at the LHC}

Due to the $\Z_2$ symmetry the inert scalars are produced in pairs. The dominant production processes at the LHC are $\hpm H$, $\hpm A$ and $\hp \hm$. The heavier scalar $\hpm$ ($H$)  eventually decays into the SM gauge boson $W$ ($Z$) and the stable inert scalar $A$ that escapes the detector giving rise to $\met$.  Depending on the decay modes of $W,Z$, various final states can be observed at future LHC experiments(e.g. jets $+ \met$, leptons $+ \met$, jets $+$ leptons $+ \met$). For reasons discussed in the introduction we focus on the $m$-leptons +  $\met$ + $X$ topologies with $m =$ 3, 4 and 5, where $X$ stands for any number of jets. Thus if some of the gauge bosons in the final state provide the required number of leptons, the others decay hadronically.

In our analysis, we have used {\tt FeynRules} \cite{fr} to generate IDM model files and {\tt micrOMEGAs} \cite{micro} to calculate relic density of $A$. Signal events are generated using {\tt CALCHEP 3.6.23} \cite{calc} and hadronization, showering are done by {\tt PYTHIA 6.4} \cite{pythia} using calchep-pythia interface. Each Background event is generated with one extra jet at parton level using ALPGEN \cite{alpgen} with MLM matching \cite{mlm} scheme to avoid double counting of jets and then passed to {\tt PYTHIA} for hadronization and showering. Jets are reconstructed using Fastjet \cite{fastjet} with anti-$K_T$ \cite{antikt} algorithm using the size parameter $R = 0.5$ and jet $P_T$ threshold of 20 GeV and $|\eta| < 2.5$. Each lepton is selected with $P_T > 10$ GeV, $|\eta| < 2.5$ and is required to pass the isolation cuts as defined by the ATLAS and CMS Collaborations \cite{atlas3l,cms3l}. We have used CTEQ6L \cite{cteq6l} for Parton Distribution Functions for all our simulations.

\subsection{The $3l + \met$ signal}

First, we concentrate on the $3l + \met +X$ final state. The dominant contribution comes from the 
processes : 
\begin{itemize}
\item $p p \ra \hpm H$ followed by $\hpm \ra W^{\pm} A / W^{\pm} H $, $H \ra Z A $  
\end{itemize}
However, depending on $\delhpmh = |\mhpm - \mh|$ the following processses may also contribute  
\begin{itemize}
\item $p p \ra \hpm A$ followed by $\hpm \ra W^{\pm} H$, $H \ra Z A $      
\item $p p \ra \hpm \hmp$ followed by $\hpm \ra W^{\pm} H / W^{\pm} A $    
\end{itemize}
$W^{\pm}$ and $Z$ decay into leptons ($W^{\pm} \ra l^\pm \nu$, $Z \ra l^+ l^-$), where $l$ stands for e or $\mu$. Here $\hpm$ decays to $W^{\pm} A$ with Branching Ratio (BR) almost $100\%$ for $\mhpm > (M_{W} + \mdm)$. For lower $\mhpm$, the decay can occur via a virtual $W^{\pm}$ boson. $\hpm$ can also decay into $H W^{\pm}$ with sufficiently large BRs if the decay is kinematically allowed. In most cases $H$ will decays into
$Z A$ with $100\%$ BR, where $Z$ can be either off-shell or on-shell depending on $\delh$.
However, if $\mh > \mhpm$, $H \ra W^{\mp} \hpm$ may be a competing mode.

The dominant SM backgrounds giving trilepton final states are : 
\begin{itemize}
\item $W^\pm Z$ production where both $W^\pm$ and $Z$ decay into leptons. 
\item $Z Z$ production followed by leptonic decays of both $Z$ bosons.
\item $t \bar{t} Z$ followed by $Z \ra l^+ l^-, t(\bar{t}) \ra b(\bar{b}) W^+ (W^-)$ and one of the two $W$'s decays into leptons.
\item $VVV$ (where $V=W^\pm,Z$) production where leptonic decays of $W^\pm,Z$ may lead to final states with $3l + \met$. 

\end{itemize}

In this paper we focus on the experiments at 13 TeV. In order to suppress the large SM background we have employed the following cuts $\colon$

\begin{itemize}

\item Exactly 3 isolated leptons are required.
\item $\met > 100$ GeV.
\item If invariant mass of any SFOS (Same Flavour Opposite Sign) pair of leptons is found to be in the range  81.2--101.2 GeV, the event is rejected.
\end{itemize}

The dominant WZ background inevitably contains an on-shell Z boson, i.e., the invariant mass distribution of a SFOS lepton pair peaks around $M_Z$. This immediately suggests that the last cut can suppress the background significantly if the signal events do not contain an on-shell $Z$ boson. The number of background events after all cuts is $4039.8$ for an integrated luminosity of 3000 fb$^{-1}$.


We have introduced in section 3 three representative scenarios A) - C). Benchmark points (BP1 - BP12) satisfying all constraints discussed in section 3 are chosen from these scenarios and displayed in Table \ref{tab1}. The relevant production cross sections and BRs are listed in Table \ref{tab2}. From this information it readily follows that the event rates for different multilepton signals before applying the kinematical selections is already modest. Moreover the spectrum of the additional scalars are somewhat compressed due to the constraints discussed in the last section.
As a result the $\met$ spectra of various signals tend to be rather soft.
This is why one has to wait for sufficiently large integrated luminosity for observing the signals as we shall see in this section.

We begin with case B) (Fig.~2 of \cite{Khan:2015ipa}) with $\lambda_S = 0.007$ and $M_A=70$ GeV (see BP1 - BP4).  If we require unitarity  upto the Planck scale (i.e., a grand desert type scenario) a tiny part of the total parameter space (the white region in the lower left corner of Fig.~2 of \cite{Khan:2015ipa}) survive. The intersection of this region with the parameter space allowed by the DM  (the upper and lower blue bands), the $T$-parameter and other constraints constitute the APS. In the entire APS both $W$ and $Z$ bosons in the signal  are off-shell since $\delhpm$ and $\delh$ are relatively small. This, as discussed above,  enables one to  probe  the whole APS via the trilepton signal  at the LHC with an integrated luminosity of $\approx$ 3000 ${\rm fb^{-1}}$. The  significance of the signal for BP1 belonging to the lower narrow blue band (see Table 3) is encouraging. Similar promising results for BP2 and BP3 in the upper blue band  are also in the same Table.

As already noted in section 3 if the scale of the validity of the unitarity constraint is relaxed to 10 TeV assuming this to be the onset of some new physics the entire parameter space shown in Fig. 2 of \cite{Khan:2015ipa} becomes consistent with this relaxed constraint. Consequently the entire broader blue band, subject to the $T$-parameter and other constraints, belongs to the APS. However, only points with $\delh \leq M_Z$ can be probed at the LHC as has already been noted. On the other hand almost all $\mhpm$ allowed by the vertical width of the APS can be probed by the future LHC experiments with integrated luminosity 3000 fb$^{-1}$. In Table \ref{tab3} BP4 near the upper edge of the APS illustrates the significance of the signal.

In scenario A) represented by BP5-BP8,  the upper blue band allowed by the DM constraint is shifted to higher $\delhpm$ compared to that in scenario B) (see Fig.~\ref{fig:STUcheck}). As a result the  stronger unitarity, the DM and the $T$-parameter constraints  allow  a small APS consisting of a part of the horizontal blue band corresponding to $\delhpm \approx 15$ GeV. Similar conclusions hold for still smaller  $\lambda_S$ as can be seen from   Fig.~5 of \cite{Khan:2015ipa} with $\lambda_S=0.001$. In all such cases the entire APS  can be  probed by the LHC experiments at 13 TeV with integrated luminosity of about 3000 fb$^{-1}$. 
If instead the weaker unitarity constraint is invoked  the APS includes a part of the broader blue region in Fig.~\ref{fig:STUcheck} consistent with all constraints. The portion corresponding to $\delh < M_Z$ can be probed as is illustrated by    
BP5-8 in Table \ref{tab3}. Note that for BP6 - 8, $\delhpmh$ is  larger than $M_W$ so that the decay  $\hpm \rightarrow H W$ is kinematically allowed and occurs  with fairly large BR (see Table~\ref{tab2}). As a result,  the pair production processes   $\hpm H$, $\hpm A$ and $\hp \hm$ can  contribute to $3l + \met + X$ signal. These processes are also potential sources of the $4l$ and $5l$ signatures  which will be discussed  below. It can be readily seen from Fig.~\ref{fig:STUcheck} that  for $\delhpm \gtrsim 210$ GeV, the $T$  parameter constraint implies that  $H$ necessarily decays into  on-shell Z bosons. Thus in this scenario $\mhpm  \leq$ 280 GeV can be probed by the future LHC experiments for the above integrated luminosity. This is illustrated by BP8 in table III. As discussed above lowering $\lambda_S$  shifts the broader blue band upwards.  As a result  the part of the APS accessible to  LHC will further shrink for smaller $\lambda_S$.

Next we discuss the scenario C) with $M_A = 55$ GeV and $\lambda_S = 0.0035$. In the APS compatible with  the strong unitarity condition, the LEP and the DM constraints (see Fig.~\ref{fig:STUcheck55}) we have $\delh < M_Z$. Thus the entire allowed region can be probed with integrated luminosity 3000 $fb^{-1}$. If instead the relaxed unitarity is imposed then  the APS is larger   
(see Fig.~\ref{fig:STUcheck55}). If $\delh < M_Z$ is required for a healthy signal,  $\delhpm$ must be in the range $12-210$ GeV. The BPs 9-12 in Table~\ref{tab1} illustrate this. For BPs 9 and 10, $\hpm$ is much heavier than $H$ leading to the decay  $\hpm \rightarrow  W H$ with appreciable BR. On the other hand, for BP11 $H$ is heavier its decay into $\hpm W$ with a BR  large enough for a multilepton signal is allowed. BP12 represents the reach in  $M_{\hpm}$. This scenario can potentially lead to the invisible Higgs decay signal. The additional confirmation may indeed come from the multilepton signatures.

\begin{table}[h]
\begin{center}
\begin{tabular}{|c||c|c|c|c|}

\hline
\hline
Benchmark  &$\lambda_S$ &\multicolumn{3}{c|}{Masses in GeV} \\
\cline{3-5} 
Points & &$\mdm$ &$\mhpm$ &$\mh$     \\
\cline{1-5}

\hline
\hline
BP1 & & &85  &140 \\
BP2 &0.007 &70 &120 &150 \\
BP3 & & &150 &140 \\
BP4 & & &170 &120 \\
\hline
BP5 & & &200 &150      \\
BP6 & 0.005& 70&240 &130  \\
BP7 & & &260 &120    \\
BP8 & & &280 &160    \\
\hline
BP9 & & &75 &135  \\
BP10 &0.0035 &55 &175 &125 \\
BP11 & & &235 &115 \\
BP12 & & &265 &115 \\
\hline
\hline

\end{tabular}
\end{center}
\caption{ A list of the BPs used in our analysis.}
\label{tab1}
\end{table} 

\begin{table}[h]
\begin{center}
\begin{tabular}{|c||c|c|c|c|c|c|}

\hline
\hline
Cross-sections &\multicolumn{6}{c|}{Benchmark points} \\
\cline{2-7}
and BRs &BP1 &BP2 &BP3 &BP4 &BP5 &BP6 \\
\cline{1-7}
$\hpm H$ &235.8 &124.7 &96.73 &94.97 &47.72 &37.09 \\
$\hpm A$ &- &- &- &- &- &61.78 \\
$\hp \hm$ &- &- &- &- &- &10.07 \\ 
\hline
$\hpm \ra l^{\pm}\nu A$ &0.226 &0.226 &0.225 &0.224 &0.226 &0.188 \\
$\hpm \ra l^{\pm} \nu H$ &* &* &* &* &* &0.042 \\
$H \ra l^+ l^- A$ &0.037 &0.067 &0.069 &0.069 &0.069 &0.069 \\
$H \ra l^{\pm} \nu \hmp$ &0.101 &* &* &* &* &* \\
\hline
\hline
Cross-sections &\multicolumn{6}{c|}{Benchmark points} \\
\cline{2-7}
and BRs &BP7 &BP8 &BP9 &BP10 &BP11 &BP12 \\
\cline{1-7}
$\hpm H$ &32.49 &19.42 &298.7 &84.0 &44.76 &32.05 \\
$\hpm A$ &47.95 &37.70 &- &- &74.31 &49.91 \\
$\hp \hm$ &7.72 &6.07 &- &- &10.94 &7.35 \\ 
\hline
$\hpm \ra l^{\pm}\nu A$ &0.15 &0.183 &0.225 &0.224 &0.168 &0.148 \\
$\hpm \ra l^{\pm} \nu H$ &0.070 &0.036 &* &* &0.052 &0.071 \\
$H \ra l^+ l^- A$ &0.069 &0.069 &0.042 &0.069 &0.069 &0.069 \\
$H \ra l^{\pm} \nu \hmp$ &* &* &0.044 &* &* &* \\
\hline
\hline

\end{tabular}
\end{center}
\caption{Leading order cross-sections (in fb) for $\hpm H$, $\hpm A$ and $\hp \hm$ production processes at $\sqrt{s} = 13$ TeV and their leptonic BRs for BPs defined in Table \ref{tab1}. For each BP only the processes that can lead to $ \geq 3$ leptons in the final state are shown. `-' indicates that the process cannot give multileptons and `*' indicates that the corresponding decay mode is absent.}
\label{tab2}
\end{table}

In Table \ref{tab3}, we summarize our results for $3l + \met$ final state assuming an integrated luminosity of 3000 fb$^{-1}$ where the total SM background ($B$) is found to be 4039.8. For almost all the BPs, the significance of the signal exceeds $5\sigma$ which indicates a good chance of discovery in future LHC experiments. The numbers in brackets indicate the signal significance for 300 fb$^{-1}$ of integrated luminosity. Note that, some of them highlights early hints of new physics at the LHC.
\begin{table}[h]
\begin{center}
\begin{tabular}{|c|c|c||c|c|c|}

\hline
\hline
Benchmark &Signal events &$S/\sqrt{B}$ &Benchmark &Signal events &$S/\sqrt{B}$  \\
Points  &after cuts ($S$) &  &Points &after cuts ($S$) & \\
\hline
BP1 &405.0 &6.37(2.01) &BP7 &502.3 &7.90(2.50) \\
BP2 &535.8 &8.42(2.66) &BP8 &373.7 &5.87(1.85) \\
BP3 &442.2 &6.95(2.19) &BP9 &641.5 &10.1(3.19) \\
BP4 &298.8 &4.70(1.48) &BP10 &491.6 &7.73(2.44) \\
BP5 &372.0 &5.85(1.85) &BP11 &676.6 &10.64(3.36)  \\
BP6 &472.8 &7.43(2.35) &BP12 &617.8 &9.72(3.07) \\
\hline
\hline

\end{tabular}
\end{center}
\caption{The number of $3l$ events (denoted by $S$) surviving the cuts defined in the text and also the statistical significance (defined as ${ S/\sqrt{B}}$) at $\sqrt{s} =13$ TeV for an integrated luminosity of 3000 fb$^{-1}$ are given for all BPs defined in Table \ref{tab1}. The number of total SM background (denoted by $B$) is 4039.8. The numbers in brackets represent the statistical significance for 300 fb$^{-1}$ of integrated luminosity.}
\label{tab3}
\end{table}


\subsection{$4l + \met$ signal}

In this section, we discuss the $4l + \met +X$ signatures. The dominant contributions to signal comes from the processes : \\
\begin{itemize}
\item $p p \ra \hpm H$ followed by $\hpm \ra W^{\pm} H$ and $H \ra l^+ l^- A$  
\item $p p \ra \hp \hm$ where either both $\hp$ and $\hm$ decay into $W^+ H$ or one into $W^- H$ and other into $W^\pm A$ 
\end{itemize} 

The main SM backgrounds are the following :
\begin{itemize}

\item $Z Z$ production followed by the leptonic decays of both $Z$ bosons, $Z \ra l^+ l^-$.
\item $W^\pm W^\mp Z$ production where both $W^\pm$ and $Z$ decay leptonically, $W^\pm \ra l^\pm \nu$ and $Z \ra l^+ l^-$
\item $W^\pm Z Z$ production where two $Z$ bosons decay into lepton pairs.
\item $Z Z Z$ production followed by leptonic decays of any two $Z$ bosons.
\item Finally $t \bar{t} Z$ production followed by $Z \ra l^+ l^-$ and $t(\bar{t}) \ra b(\bar{b}) W^+ (W^-) , W^\pm \ra l^\pm \nu$

\end{itemize}

We apply the following set of cuts in our analysis to suppress the background as well as to select signal events :

\begin{itemize}

\item Exactly 4 isolated leptons are required.
\item A cut of $80$ GeV on $\met$. Note that, this cut is strong enough to suppress the potentially strong background coming from $Z Z$ which has a comparatively soft $\met$ distribution.
\item Finally, an event with at least one SFOS lepton pair with invariant mass in the range $81.2-101.2$ GeV is rejected.  

\end{itemize}

After applying the above cuts, the number of total background events ($B$) reduces to 36.9 for 3000 fb$^{-1}$ of integrated luminosity. 

Table \ref{tab4} shows the simulation results corresponding to the BPs introduced in Table \ref{tab1}. In many cases observable signals  with integrated luminosity somewhat larger than the typical choice of 3000 fb$^{-1}$ may be expected.

\begin{table}[h]
\begin{center}
\begin{tabular}{|c|c|c||c|c|c|}

\hline
Benchmark &Signal events & $S/\sqrt{B}$ &Benchmark &Signal events &$S/\sqrt{B}$  \\
Points   &after cuts ($S$) &  &Points &after cuts ($S$) &\\
\hline
BP6 &21.49 &3.53 &BP11 &31.83 &5.30 \\
BP7 &21.16 &3.45 &BP12 &27.65 &4.55 \\
BP8 &9.79 &1.59 & & & \\
\hline

\end{tabular}
\end{center}
\caption{Number of $4l$ events ($S$) along with the statistical significances at $\sqrt{s} = 13$ TeV for an integrated luminosity of 3000 fb$^{-1}$. BPs are taken from Table \ref{tab1}. The number of SM background ($B$) is 36.9.}
\label{tab4}
\end{table}

\subsection{$5l + \met$ signal}

In this section, we examine the prospects for the $5l + \met+X$ signal at future LHC experiments. The main process contributing to the signal is 
\begin{center}

         $p p \ra \hpm H$ followed by $\hpm \ra W^\pm H, W^\pm \ra l^\pm \nu$ and $H \ra l^+ l^- A$ \\
\end{center}

Note that $\hp$ pair production (where both $\hp$ and $\hm$ decay into $W H$), in principle, can also give $5l$ final states. But we have found this contribution to be negligible. We list below the SM backgrounds. 

\begin{itemize}

\item $Z Z Z$ production followed by leptonic decays of all $Z$ bosons, where one lepton is not detected or fails to pass the cuts.
\item $W^\pm Z Z$ production with both $W^\pm$ and $Z$ decaying into leptons.
\item $t \bar{t} Z$ production where the corresponding decay occurs via $Z \ra l^+ l^-$, $t(\bar{t}) \ra b(\bar{b}) W^+ (W^-) , W^\pm \ra l^\pm \nu$ and one lepton comes from $b$ decay ($b \ra c l \nu$)
\end{itemize}

Demanding $5$ isolated leptons in the final state drastically reduces the background. A cut of $80$ GeV on $\met$ is good enough to efficiently reduce the background to a negligible level. As a rough guideline we require for discovery at least five background free events.  We present the  results in Table \ref{tab5}.

\begin{table}[h]
\begin{center}
\begin{tabular}{|c|c||c|c|}

\hline
Benchmark  &Signal events  &Benchmark &Signal events  \\
Points   &after cuts ($S$)   &Points &after cuts ($S$) \\
\hline
BP6 &3.33  &BP11 &5.37  \\
BP7 &3.89  &BP12 &1.92  \\
BP8 &1.75  & &  \\

\hline

\end{tabular}
\end{center}
\caption{Number of $5l$ events ($S$) at $\sqrt{s} = 13$ TeV for an integrated luminosity of 3000 fb$^{-1}$ for the BPs defined in Table \ref{tab1}. The SM background is negligible.}
\label{tab5}
\end{table}

As we  see from Table \ref{tab5}, when BP11 has potential discovery chance at $\sqrt{s} = 13$ TeV, rest of the BPs may provide hints for the IDM at LHC until  higher luminosities well beyond $ 3000 fb^{-1}$ accumulate. It also follows from Tables III - V that the relative rates of different signals can discriminate among different IDM scenarios.

\section{Conclusion}

The aim of this paper is to revisit the prospect of observing the $m l +\met + X$ signatures predicted by the IDM, a popular DM model,  in future LHC experiments for $m =$ 3, 4. It may be recalled that the earlier studies \cite{ Miao:2010rg,Gustafsson:2012aj} were based on BPs disfavoured by  the strong LHC constraints in the post Higgs discovery era. In this context the accurate measurement of the Higgs boson mass and the stringent upperbound on the invisible width of the Higgs boson deserve special mentioning. We also simulate for the first time the $5l + \met$ signal and study its observability.

To facilitate our analyses we introduce at the beginning of Sec. 3 three representative scenarios A), B) and C) and delineate the APS in each case subject to the constraints discussed in the same section (see Figs.~\ref{fig:STUcheck} and \ref{fig:STUcheck55} and Fig. 2 of \cite{Khan:2015ipa}). Following the search strategies in section 4 we then assess the prospect of discovery of each signal. As discussed in this section the signals are viable only if the leptons come from the virtual $Z$ bosons (i.e., $\Delta M_H = M_H-M_A < M_Z$). If the IDM is embedded in a grand dessert type scenario, i.e., the unitarity constraint is required to be valid upto, e.g., the Planck scale, then in each scenario the APS is tiny with $\Delta M_H < M_Z$. Thus the entire allowed parameter space in all scenarios can be probed via the $3l$ signal for integrated luminosity $\sim$ 3000 fb$^{-1}$. Although one has to wait for the LHC experiments after the third long shut down to achieve this, these results shows that the grand dessert type IDM models are definitely falsifiable.

If the unitarity constraint is relaxed to a lower scale, the APS, as expected, is larger in each case. For $\Lambda = 10$ TeV, we have shown that the entire APS (i.e., the allowed range of $\Delta{M_{H^{\pm}}}$ with $\Delta M_H < M_Z$) are  accessible to the LHC experiments in all three scenarios with $\sim$ 3000 ${\rm fb^{-1}}$ of integrated luminosity. We also point out  that the accessible region shrinks for smaller $\lambda_S$ due to the  strong $T$-parameter constraint.

The above observations are substantiated by numerical results (see Tables III - V) for the BPs in Table I. The relative rates of different $ml$ signals can discriminate among different scenarios.


\vskip 20pt
\noindent{\bf Acknowledgements:}\\
The research of A.D. was supported by the Indian National Science Academy, New Delhi. N.G. thanks the Board of Research in Nuclear Sciences, Department of Atomic Energy, India for a research fellowship. The work of N.K. is supported by a fellowship from University Grants
Commission, India. S.R. is funded by the Department of Science and Technology, India via Grant No. EMR/2014/001177.  S.R. acknowledges Siddhartha Karmakar for discussion.


\end{document}